\documentclass[12pt]{article}
\usepackage{amssymb}
\usepackage[dvips]{epsfig}

\newcommand{\be}{\begin{equation}}
\newcommand{\ee}{\end{equation}}
\newcommand{\bn}{\begin{eqnarray}}
\newcommand{\en}{\end{eqnarray}}

\newcommand{\nn}{\nonumber}
\newcommand{\emn}{\epsilon_{\mu\nu\gamma}\partial^{\gamma}}

\newcommand{\no}{\noindent}

\def\bea{\begin{eqnarray}}
\def\eea{\end{eqnarray}}

\def \ps{\partial\!\!\!/}
\def \fs{f\!\!\!/}
\def \sd{D\!\!\!/}
\newcommand{\beq}{\begin{eqnarray}}
\newcommand{\eeq}{\end{eqnarray}}
\begin{document}

\title{\textbf{Ghost free dual vector theories in $2+1$ dimensions }}
\author{D. Dalmazi \\
\textit{{UNESP - Campus de Guaratinguet\'a - DFQ} }\\
\textit{{Av. Dr. Ariberto Pereira da Cunha, 333} }\\
\textit{{CEP 12516-410 - Guaratinguet\'a - SP - Brazil.} }\\
\textsf{E-mail: dalmazi@feg.unesp.br }}
\date{\today}
\maketitle

\begin{abstract}
We explore here the issue of duality versus spectrum equivalence in abelian vector
theories in $2+1$ dimensions. Specifically we examine a generalized self-dual (GSD)
model where a Maxwell term is added to the self-dual model.  A gauge embedding
procedure applied to the GSD model leads to a Maxwell-Chern-Simons (MCS) theory with
higher derivatives. We show that the latter contains a ghost mode contrary to the
original GSD model. On the other hand, the same embedding procedure can be applied to
$N_f$ fermions minimally coupled to the self-dual model. The dual theory corresponds
to $N_f$ fermions with an extra Thirring term coupled to the gauge field via a
Pauli-like term. By integrating over the fermions at $N_f\to\infty$ in both matter
coupled theories we obtain effective quadratic theories for the corresponding vector
fields. On one hand, we have a nonlocal type of the GSD model. On the other hand, we
have a nonlocal form of the MCS theory. It turns out that both theories have the same
spectrum and are ghost free. By figuring out why we do not have ghosts in this case we
are able to suggest a new master action which takes us from the local GSD to a
nonlocal MCS model with the same spectrum of the original GSD model and ghost free.
Furthermore, there is a dual map between both theories at classical level which
survives quantum correlation functions up to contact terms. The remarks made here may
be relevant for other applications of the master action approach.

 \textit{{PACS-No.:}
11.15.-q, 11.10.Kk, 11.10.Gh, 11.10.Ef }
\end{abstract}

\newpage

\section{Introduction}

The bosonization program in two dimensions is a good example of the power of dual
descriptions of the same theory \cite{Coleman,Mandelstam}, see also \cite{Abdalla} and
references therein. Although much less progress has been done in this area in higher
dimensions, some nonperturbative features like confinement can also be revealed with
the help of duality  as in \cite{SW}. Here we are interested in the intermediate case
of $2+1$ dimensions. By means of a master action it was shown in \cite{DJ} that the
gauge invariant sector of the Maxwell Chern-Simons (MCS) theory is on shell equivalent
to the self-dual (SD) theory of \cite{TPN} with the dual map $f_{\mu}\leftrightarrow
\frac{\epsilon_{\mu\nu\alpha}\partial^{\nu}A^{\alpha}}m$. This type of duality is
intimately related to the bosonization program in $d=2+1$
\cite{FS,Kondo,Banerjee1,Barci,ddh2,BottaHelayel}.

The  master action of \cite{DJ} can be generalized to include matter fields as in
\cite{GMS} and \cite{Anacleto2,jpa}. Although very useful, specially at quantum level,
the master action is not derived from first principles and it is justified {\it a
posteriori}. A systematic derivation of dual gauge theories would certainly be
welcome. In \cite{Anacleto2} a Noether gauge embedding procedure was suggested and
applied in a variety of examples and dimensions \cite{Menezes,Bazeia,Bazeia2}.
However, it has been pointed out in \cite{Baeta1} that this procedure may lead to
ghosts in the dual gauge theory which has been explicitly verified in a dual model to
a four dimensional Lorentz violating electrodynamics \cite{Baeta2}. An understanding
of this issue from the point of view of master actions allows us to propose an
alternative master action to avoid the problem which is the aim of this work. We start
in the next section from a rather general quadratic theory for a vector field and
calculate the residue of its propagator around its poles in order to verify the
presence of ghost modes. The so called Noether embedding procedure is briefly reviewed
as well as its use to generate a dual gauge theory. In section 3 we compare the
spectrum of a GSD theory of non-local type and its dual gauge model. Both theories are
generated by a $N_f \to\infty$ limit of parity invariant fermions ($4\times 4$
representation) coupled to vector fields. The equivalence of both spectra in this case
leads us to formulate a new master action in section 4 to solve the problem of ghosts
in the dual theory. In the last section we draw some conclusions.

% but it is also at the heart of
%another important issue in field theory which goes beyond the $2+1$ dimensional world,
%namely, the definition of local and sensible theories for a massive gauge field.

\section{The GSD model and its higher derivative dual}

\no Allowing arbitrary functions of the D'Alambertian $a_i=a_i(\Box)$, the Lagrangian
below includes a rather general set of Poincar\'e covariant quadratic theories for an
abelian vector field in $d=2+1$, see also \cite{Oswaldo,Bazeia2},

\be {\cal L}_{GSD} \, = \, a_0 f^{\mu}f_{\mu} + a_1
\epsilon_{\alpha\beta\gamma}f^{\alpha}\partial^{\beta}f^{\gamma} + \frac{a_2}2
F_{\mu\nu}(f)F^{\mu\nu}(f) \label{gsd} \ee

\no We use $g_{\mu\nu}=(+,-,-)$ and  assume in this work that $a_i$ are constants but
later on we will make a comment on the general case. Minimizing ${\cal L}_{GSD}$ we
can write its equations of motion in a self-dual form where $\tilde{f}$ defined below
stands for the dual field:

\be f_{\mu} \, = \, \frac{a_1}{a_0} E_{\mu\nu} f^{\nu} + \frac{a_2}{a_0} \Box \,
\theta_{\mu\nu} f^{\nu} \, \equiv \, {\tilde f}_{\mu} \label{sdeq} \ee

\no where $E_{\mu\nu} = \emn $ and $\Box \,\theta_{\mu\nu} = g_{\mu\nu} -
\partial_{\mu}\partial_{\nu}$. From (\ref{sdeq}) we get $\partial_{\mu}f^{\mu}=0$.
Since $E_{\mu\nu}E^{\nu\gamma} = - \Box \, \theta_{\mu}^{\gamma} $ and both $a_0$ and
$a_1$ are assumed to be non-vanishing, it is easy to derive from (\ref{sdeq}):

\be \left\lbrack \left(a_0 - a_2 \Box \right)^2  + a_1^2 \Box \right\rbrack f^{\mu} \,
= \, a_2^2 \left( \Box + m_+^2\right)\left( \Box + m_-^2\right)f^{\mu} \, = 0
\label{4th} \ee

\no with

\be 2 m_{\pm}^2 \, =\, b^2 - 2 a \pm \sqrt{(b^2-2a)^2-4 a^2} \label{mpm} \ee

\no and  $a=a_0/a_2 \, , \, b=a_1/a_2$. The masses are real for $a < 0$ or $b^2 > 4a$
if $a > 0 $. An interesting observation is that we can not have $\left(a_0 - a_2 \Box
\right)f^{\mu} = 0 $, since from (\ref{4th}) we see that this would lead to $\Box
f^{\mu}=0$ and this two equations together would contradict our hypothesis $a_0 \ne
0$. We can read off the propagator for the generalized self-dual field  from
(\ref{gsd}). In the momentum space we have:

\be \left\langle f_{\alpha}(k) f_{\beta}(-k) \right\rangle_{GSD} \, = \,
\frac{g_{\alpha\beta}-\theta_{\alpha\beta}}{a_0} + \frac{\left(a_0 + k^2
a_2\right)}{\left(a_0 + k^2 a_2\right)^2 - a_1^2 k^2} \theta_{\alpha\beta} + \frac{i\,
a_1 E_{\alpha\beta}}{\left(a_0 + k^2 a_2\right)^2 - a_1^2 k^2} \label{pgsd} \ee

\no As expected from (\ref{4th}) there are two simple poles at $k^2=m_{\pm}^2$ which
become a double pole only at the special point $a_1^2 = 4 a_0 a_2 $. In order to make
sure that we are free of ghosts we have to show in general, see e.g.
\cite{Baeta2,Oswaldo,DN}, that the imaginary part of the residue of the propagator at
each pole is positive when saturated with conserved currents, that is
$Im\left(Res(J_{\alpha}\left\langle f^{\alpha}(k) f^{\beta}(-k)\right\rangle
J^*_{\beta} )\right) > 0 $ with $k_{\mu}J^{\mu} =0=k^{\mu}J_{\mu}^*$. In $d=2+1$ the
propagator of a vector field in a Poincar\`e covariant theory will have the general
Lorentz structure $A(k^2)(g_{\mu\nu}-\theta_{\mu\nu}) + B(k^2) \theta_{\mu\nu} +
C(k^2) E_{\mu\nu}$. In practice\footnote{By choosing a convenient frame for a massive
pole $k=(k,0,0)$, which implies $J_{\mu}=(0,J_1,J_2)$, and using the general formula
for a propagator in $d=2+1$ one can show that  $Im\left(Res(J_{\alpha}\left\langle
f^{\alpha}(k) f^{\beta}(-k)\right\rangle J^*_{\beta} )\right) = - S J\bar J = - S (J_1
+ i J_2)(J_1^*-i J_2^*) $ where for every massive single pole $S=\lim_{k^2\to m^2}(k^2
- m^2)B(k^2)$.}, this is equivalent to require $\lim_{k^2\to m^2}(k^2 - m^2)B(k^2) < 0
$. Applying this requirement at $k^2=m_+^2$ and $k^2=m_-^2$ respectively we have:

\beq \frac{a + m_+^2}{a_2 (m_+^2 - m_-^2)} \, &=& \, \frac{b^2 + \sqrt{b^2(b^2-4a)}}{2
a_2 (m_+^2 - m_-^2)} \, < \, 0 \label{req1} \\
\frac{a + m_-^2}{a_2 (m_-^2 - m_+^2)} \, &=& \, -\frac{b^2 - \sqrt{b^2(b^2-4a)}}{2 a_2
(m_+^2 - m_-^2)} \, < \, 0 \label{req2} \eeq

\no Since $m_+^2 > m_-^2$ we are free of ghosts if $a_0 > 0 $ and $a_2 < 0$, in
agreement with \cite{Oswaldo}. Those are in fact usual conditions on the mass term of
the self-dual model and on the coefficient of the Maxwell term in QED in $d=2+1$
respectively. Notice that we get rid automatically of the double pole under such
circumstances.

In \cite{Bazeia2} the gauge theory dual to ${\cal L}_{GSD}$ was derived via a Noether
gauge embedding procedure which in summary works as follows. Since the equations of
motion for the the self-dual field come from $K^{\mu} = 2\left( a_0 f^{\mu} - a_1
E^{\mu\nu}f_{\nu} - a_2 \Box\,\theta^{\mu\nu}f_{\nu}\right) = 0 $, and under a local
$U(1)$ transformation $\delta^{\phi}f_{\mu} =
\partial_{\mu}\phi $ we have $\delta^{\phi} K_{\mu} = 2 a_0 \delta^{\phi}f_{\mu}$.
Thus, if we define ${\cal L}_{HD-MCS}= {\cal L}_{GSD} - K_{\alpha}K^{\alpha}/4a_0$ it
follows that $\delta^{\phi} {\cal L}_{HD-MCS} = 0$ and we have a gauge invariant
theory. By minimizing this theory $\delta {\cal L}_{HD-MCS} = K^{\mu} \delta f_{\mu} -
K^{\mu} \delta K_{\mu}/2r_0 = 0 $ it is clear that the equations of motion of ${\cal
L}_{GSD}$ ($K_{\mu}=0$) lead to $\delta {\cal L}_{HD-MCS}=0$ and therefore are
embedded in the equations of motion of the new gauge theory. We stress that such
embedding does not guarantee equivalence of the equations of motion of both theories,
not even in the gauge invariant sector of ${\cal L}_{HD-MCS}$. Applying this embedding
explicitly, renaming the field $f_{\mu}\to A_{\mu} $, one has the higher derivative
theory \cite{Bazeia2} :

\beq {\cal L}_{HD-MCS}\, &=& \, {\cal L}_{GSD} - \frac{K_{\alpha}K^{\alpha}}{4a_0} \nn
\\
&=& \, A^{\mu}a_1\left(1-\frac{2 a_2\Box}{a_0}\right)\emn A^{\nu} - \frac 12
F_{\mu\nu}(A)\left(\frac{a_1^2}{a_0} - \frac{a_2^2\Box}{a_0} + a_2\right)F^{\mu\nu}(A)
\label{gmcs} \eeq

\no The equations of motion of the higher derivative model (\ref{gmcs}) can be written
as

\be  a_1\left(1-\frac{2 a_2\Box}{a_0}\right)\emn A^{\nu} + \Box
\left(\frac{a_1^2}{a_0} - \frac{a_2^2\Box}{a_0} + a_2\right)\theta_{\mu\nu}A^{\nu} \,
= a_0 \left(\tilde{A}_{\mu} - \tilde{\tilde{A}}_{\mu}\right) \, = \, 0 \qquad ,
\label{gmcseq}\ee

\no which makes evident, see (\ref{sdeq}), the duality at classical level with the GSD
model upon replacing $f_{\mu}$ by $\tilde{A}_{\mu}$. From (\ref{gmcseq}) we deduce the
analogous of formula (\ref{4th}):

\be \left(a_1^2 + a_2^2\, \Box \right)\left(\Box + m_+^2\right)\left(\Box +
m_-^2\right)\emn A^{\nu} \, = \, 0 \label{6th} \ee

\no The expression (\ref{6th}) shows that we have new classical solutions in the gauge
invariant sector $\left(a_1^2 + a_2^2\, \Box \right)\emn A^{\nu} = 0 \, $ which were
not present in the GSD model. The new solutions will correspond to a new pole in the
propagator. After introducing a gauge fixing term we have the propagator:

\be \left\langle A_{\alpha}(k) A_{\beta}(-k) \right\rangle_{HD-MCS} \, = \,
\frac{g_{\alpha\beta}-\theta_{\alpha\beta}}{\lambda k^2} + B(k^2) \theta_{\alpha\beta}
+ \frac{i a_0 a_1 (a_0 + 2 a_2 k^2)E_{\alpha\beta}}{k^2 (a_1^2 - k^2 a_2^2
)\left\lbrack\left(a_0 + k^2 a_2\right)^2 - a_1^2 k^2\right\rbrack }\label{pgmcs0} \ee

\no where, in agreement with the general predictions of \cite{Baeta1}, we have

\be B(k^2) \, = \, \frac{\left(a_0 + k^2 a_2\right)}{\left(a_0 + k^2 a_2\right)^2 -
a_1^2 k^2} \, - \,  \frac{a_2}{a_2^2 k^2 - a_1^2} \label{bk2} \ee

\no Since the first term in $B(k^2)$ is the same one appearing in (\ref{pgsd})  it is
clear that $k^2 = m_{\pm}^2$ are still physical poles as in the GSD model for $a_0 >
0$ and $a_2 < 0 $. However, since $\lim_{k^2\to (a_1/a_2)^2} \left\lbrack k^2 -
(a_1/a_2)^2\right\rbrack B(k^2) = -1/a_2 $ should be negative the new pole will be a
ghost if we insist in $a_2 < 0 $. On the other hand, if we drop that condition one of
the poles $k^2=m_{\pm}^2$ will become a ghost mode. From (\ref{bk2}) we see that the
only possible exit might be a fine tuning of $a_0,a_1,a_2$ such that one of the poles
$m_{\pm}^2$ coincides with $(a_1/a_2)^2$ and  the corresponding residue has the right
sign. This is only possible for $a_1^2 = - (a_0 a_2)/2 $. In this case
$m_-^2=(a_1/a_2)^2$ but $B(k^2)=-a_0/\left\lbrack a_2^2 (k^2-m_-^2)(k^2-m_+^2)
\right\rbrack $ which implies that one of the poles $m_{\pm}^2$ must be a ghost again.
In conclusion, the dual gauge model ${\cal L}_{HD-MCS}$ obtained by the Noether
embedding procedure will always contain a ghost in the spectrum as far as $a_2 \ne 0$.
We only recover a ghost free theory in the well known case of the SD/MCS duality
\cite{DJ} where $a_2=0$.  For future use we notice that in \cite{Bazeia2} a master
action which relates ${\cal L}_{GSD} $ and ${\cal L}_{HD-MCS}$ was suggested:

\beq {\cal L}_{Master} \, &=& \, a_0 f^{\mu}f_{\mu} + a_1
\epsilon_{\alpha\beta\gamma}f^{\alpha}\partial^{\beta}f^{\gamma} + \frac{a_2}2
F_{\mu\nu}(f)F^{\mu\nu}(f) \nn\\
&-& a_1 \epsilon_{\alpha\beta\gamma}(A^{\alpha}-
f^{\alpha})\partial^{\beta}(A^{\gamma}-f^{\gamma}) - \frac{a_2}2
F_{\mu\nu}(A-f)F^{\mu\nu}(A-f) \label{master1} \eeq

\no The existence of a interpolating master action makes even more surprising the
equivalence of two partition functions of theories with different spectrum. In the
next section $N_f$ fermions minimally coupled to the self-dual field and its dual
theory will give us a hint on how to answer that question and avoid the ghost problem.

\section{Matter induced GSD model and its dual}

A seemingly unrelated problem is the coupling of $U(1)$ matter to the self-dual field.
In \cite{GMS} the following master partition function was suggested:

\be {\cal Z} \, = \, \int {\cal D}\bar\psi_j{\cal D}\psi_j{\cal D}A{\cal D}f \,
\exp^{i\int d^3 x {\cal L}_{Master}^{U(1)}} \label{zmaster} \ee

\no with

\beq {\cal L}_{Master}^{U(1)} \, &=& \, \frac{m^2}2 f^{\mu}f_{\mu} + \frac{m}2
f^{\mu}\emn f^{\nu} +
\bar\psi_j\left\lbrack \sd (f) - m_f \right\rbrack\psi_j  \nn\\
 &-& \frac{m}2 (A^{\mu}-f^{\mu})\emn (A^{\nu}-f^{\nu}) \label{lmasteru1} \eeq.

 \no The repeated indices are summed over ($j=1,2,\cdots , N_f$) and $ D_{\mu} (f) =
 \left(i\partial_{\mu} - e f_{\mu}/\sqrt{N_f} \right)$. Performing the Gaussian
 integral over $A^{\mu}$ we have the self-dual model minimally coupled to $N_f$
 fermionic matter fields:

 \be {\cal L}_{SD + M} \, = \, \frac{m^2}2 f^{\mu}f_{\mu} + \frac{m}2
f^{\mu}\emn f^{\nu} + \bar\psi_j\left\lbrack \sd (f) - m_f \right\rbrack\psi_j
\label{sdm}\ee

\no while integrating over $f_{\mu}$ we derive the dual gauge theory which now
includes a Pauli-like term and a Thirring term \cite{GMS} :

\bea {\cal L}_{MCS + M} = &-&\frac{1}{4}F_{\alpha\beta}(A)F^{\alpha\beta}(A) + \frac
m2
\epsilon_{\alpha\beta\gamma}A^{\alpha}\partial^{\beta}A^{\gamma} \nn \\
&-& \frac{e}{m}J_{\nu} \epsilon^{\nu\alpha\beta}\partial_{\alpha}A_{\beta} -
\frac{e^2}{2m^2}J_{\nu} J^{\nu } + \bar\psi_j\left\lbrack i\ps - m_f
\right\rbrack\psi_j \, . \label{mcsm} \eea

\no The equations of motion of (\ref{sdm}) are $f_{\mu}=\tilde{f}_{\mu}$ and $\sd
(f)\psi = 0$ while from (\ref{mcsm}) we obtain $\tilde{A}_{\mu} =
\tilde{\tilde{A}}_\mu $ and $\sd (\tilde{A})\psi = 0$ where the duality transformation
includes now the matter fields $\tilde{A}_{\mu} = -\emn A^{\nu}/m +
\left(e/(m^2\sqrt{N_f})\right)J_{\mu}$ where $J_{\mu}=\bar\psi_j\gamma_{\mu}\psi_j$.
Therefore, both theories are dual to each other at classical level through the map
$f_{\mu} \leftrightarrow \tilde{A}_{\mu}$, with the matter fields being invariant
under the dual map. In \cite{jpa} we have shown that this dual map can be extended to
the quantum level. Once again, those results do not guarantee spectrum equivalence. In
particular we have now truly interacting theories which complicates the analysis.
Notwithstanding, the integration over the fermions produces a factor $\exp\left\lbrack
N_f {\rm tr }\ln\left(i\ps -m + e\fs/\sqrt{N_f} \right)\right\rbrack $ in the
partition function (\ref{zmaster}). Due to Furry's theorem only even powers of the
vector field will survive the $1/N_f$ expansion. Neglecting the field independent
factor $ \exp\left\lbrack N_f {\rm tr}\ln\left( i\ps -m \right)\right\rbrack $ we are
left only with the quadratic term of order $e^2$ at the leading order in $1/N_f$ which
becomes exact at $N_f \to \infty$. The quadratic term is the vacuum polarization
diagram. For simplicity we assume four components fermions. By computing the
polarization tensor using a gauge invariant regularization, see e.g. \cite{prd}, we
have the effective master action at $N_f\to\infty $:

\beq {\cal L}_{Master}^{U(1)}(N_f\to\infty) \, &=& \, \frac{m^2}2 f^{\mu}f_{\mu} +
\frac{m}2 f^{\mu}\emn f^{\nu} - \frac{e^2}4 F_{\mu\nu}(f) g(\Box )F^{\mu\nu}(f)\nn\\
 &-& \frac{m}2 (A^{\mu}-f^{\mu})\emn (A^{\nu}-f^{\nu}) \label{masternf} \eeq.

\no where, defining $z=-\Box/4m_f^2 $ we have $g(\Box)=f_2/(16\pi m_f)$ with, for $0
\le z \le 1$,

\be f_2 \, = \, \left( \frac 1z - \frac{1+z}{2 z^{3/2}} \ln
\frac{1+\sqrt{z}}{1-\sqrt{z}}\right) \label{g} \ee

\no For $z < 0$ the expression for $f_2$ can be analytically continued from (\ref{g}).
We neglect the imaginary part of the action which appears above the pair creation
threshold $z > 1$. Integrating over $A_{\mu}$ we have:

\be {\cal L}_{SD + M}(N_f\to\infty) \, = \, \frac{m^2}2 f^{\mu}f_{\mu} - \frac{m}2
f^{\mu}\emn f^{\gamma} - \frac{e^2}4 F_{\mu\nu}(f) g(\Box )F^{\mu\nu}(f)
\label{sdmnf}\ee

\no which is of the GSD type with:

\be a_0 \, = \, \frac{m^2}2 \quad ; \quad a_1 \, = \, - \frac m2 \quad ; \quad a_2 \,
=\, -\frac{e^2 g(\Box)}{2} \label{aaa} \ee

\no On the other hand, integrating over $f_{\mu}$ we obtain

\be {\cal L}_{MCA + M}(N_f\to\infty) \, = \, - \frac m2 A^{\mu}\emn A^{\nu} -
\frac{m^2}4 F_{\mu\nu}(A)\frac 1{\left\lbrack m^2 + e^2 \Box\, g(\Box)\right\rbrack
}F^{\mu\nu}(A) \label{mcsmnf} \ee

\no which is {\it not} of the type (\ref{gmcs}) with the identifications (\ref{aaa}).
Therefore, the Noether gauge embedding used at tree level does not go through the
effective action at $N_f\to\infty$. Next, we check the particle content of
(\ref{sdmnf}). The propagator is of the form (\ref{pgsd}) with the identification
(\ref{aaa}). As before we only have to examine the factor:

\be \frac{\left(a_0 + k^2 a_2\right)}{\left(a_0 + k^2 a_2\right)^2 - a_1^2 k^2} \, =
\, \frac{2 r ( r + s z f_2)}{m^2 \left\lbrack ( r + s z f_2)^2 - z\right\rbrack}
\label{factor}\ee

\no with the dimensionless couplings $r\equiv m/2m_f \, ; \, s\equiv e^2/(4\pi m)$.
Since $r s
> 0$ and (\ref{factor}) is invariant under $m\to -m$ it is enough to assume $r > 0$
and $s
> 0$. First of all, we remark that for  $z < 0$ ($k^2 < 0$) the denominator of
(\ref{factor}) never vanishes since $\lim_{z\to 0} z f_2 = 0$, which implies that we
are always free of tachyons. It is a numerical result that for $0 \le z \le 1$ we
always have two simple poles. That is $D\equiv \left\lbrack ( r + s z f_2)^2 -
z\right\rbrack = (z-z_+)(z-z_-)R (z)$ where $R(z)$ never vanishes and it is finite
except at the pair creation threshold $z\to 1$ where $z f_2 \to \infty$. We have
typically the plot given in figure 1 where $z_+ > z_-$.

\begin{figure}
\begin{center}
\epsfig{figure=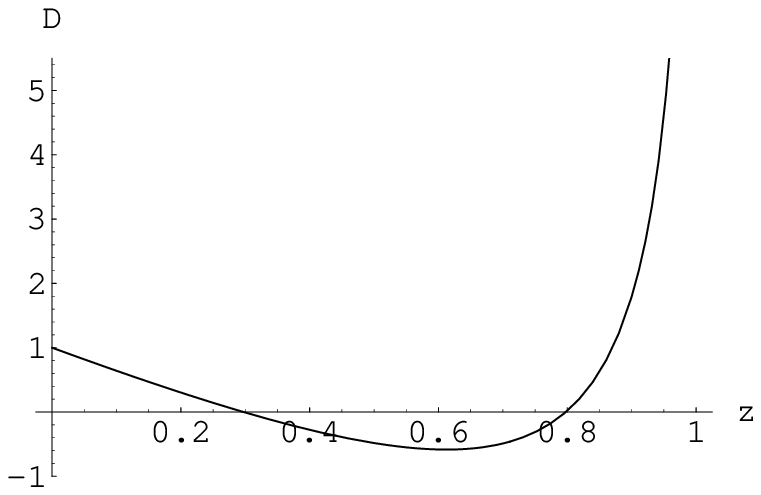,width=80mm} \caption{Plot of $D = \left\lbrack ( r + s z
f_2)^2 - z\right\rbrack$ for $r=1=s$.} \label{figure1}
\end{center}
\end{figure}

Noticing that $D = D_+ D_-$ with $D_{\pm} = \left( r + s z f_2 \pm \sqrt{z} \right)$,
we have checked, again numerically, that the poles $z=z_+$ and $z= z_-$ come
respectively from $D_+=0$ and $D_-=0$. Consequently, the sign of the residue obtained
at those poles can be written respectively in terms of the derivative ($D'$) as $\pm 2
r \sqrt{z}/\left(m^2 D'(z_{\pm})\right)$. Since , as in figure 1, $D'(z_+)
> 0$ and $D'(z_-) < 0$ both poles are physical and we are free of ghosts for arbitrary
values of $e^2 \, , m, \,  m_f $. In particular, we conclude that the interaction of
the self-dual field with the parity invariant fermions has produced, at least at
$N_f\to\infty$, an extra massive pole due to the non-local Maxwell term. Now we should
verify the spectrum of the dual gauge theory (\ref{mcsmnf}). It turns out, after
introducing a gauge fixing term, that the propagator obtained from (\ref{mcsmnf})
acquires the form:

\be \left\langle A_{\alpha}(k) A_{\beta}(-k) \right\rangle_{MCS + M} \, = \,
\frac{g_{\alpha\beta}-\theta_{\alpha\beta}}{\lambda k^2} + \frac{\left(a_0 + k^2
a_2\right)}{\left(a_0 + k^2 a_2\right)^2 - a_1^2 k^2} \theta_{\alpha\beta} + \frac{i
\left(a_0 + k^2 a_2\right)^2 E_{\alpha\beta}}{a_1 k^2 \left\lbrack \left(a_0 + k^2
a_2\right)^2 - a_1^2 k^2\right\rbrack} \label{pgmcs} \ee

\no where $a_0,a_1,a_2$ are given in (\ref{aaa}). Thus, except for the non-physical
(gauge dependent) massless pole $k^2=0$, which has a vanishing residue \cite{Baeta1},
as the reader can easily check, choosing a frame $k_{\mu}=(k,k,0)$ and saturating with
conserved currents, $\epsilon^{\mu\nu\alpha}J_{\mu}J_{\nu}^*k_{\alpha}=0$. Therefore
we have the same particle content of the ${\cal L}_{SD+M}$ model at $N_f\to\infty$. In
conclusion, the matter induced effective action preserves the particle content without
artificially creating ghost poles as in the case of the GSD/HD-MCS models in the
previous section, although the Lorentz structure of the actions is the same in both
cases.

\section{New master action for $a_2\ne 0$.}

After the translation $A_{\mu} \to A_{\mu} + f_{\mu}$ in (\ref{master1}), which has a
trivial Jacobian, we have two decoupled theories:

\be {\cal L}_{Master} \to {\cal L}_{GSD}(f) + a_1 A^{\mu}\emn A^{\nu} - \frac{a_2}2
F_{\mu\nu}(A)F^{\mu\nu}(A) \label{decoupled1} \ee

\no On the other hand, we could decouple the fields through the translation $f_{\mu}
\to f_{\mu} + \tilde{A}_{\mu}$ and obtain:

\be {\cal L}_{Master} \to {\cal L}_{HD-MCS}(A) + a_0 f^{\mu} f_{\mu}
\label{decoupled2}\ee

\no Since the spectrum of (\ref{decoupled1}) and (\ref{decoupled2}) must be the same
and $f_{\mu}$ in (\ref{decoupled2}) is certainly a non-propagating field, the poles of
${\cal L}_{GSD}(f)$ and of the MCS theory for $A_{\mu}$ in (\ref{decoupled1}) must be
contained in ${\cal L}_{HD-MCS}(A)$. Indeed, the extra pole we found in (\ref{pgmcs})
is precisely associated with the two last terms in (\ref{decoupled1}) including the
non-propagating massless pole. The above translations also explain why a master action
gives rise to two theories, in this case $ {\cal L}_{GSD}(f)$ and $ {\cal
L}_{HD-MCS}(A)$, with different spectrum.

Now it is clear why we did not have a mismatch of spectrum in the matter induced
master action. Namely, after a translation $A_{\mu} \Longrightarrow A_{\mu} + f_{\mu}$
in (\ref{masternf})we have:

\be {\cal L}_{Master}^{U(1)}(N_f\to\infty) \Longrightarrow {\cal L}_{SD
+M}(f,N_f\to\infty ) - \frac m2 A^{\mu}\emn A^{\nu} \label{decoupled3} \ee

\no If we do a convenient translation in $f_{\mu}$, with trivial Jacobian, we derive
instead:

\be {\cal L}_{Master}^{U(1)}(N_f\to\infty) \Longrightarrow {\cal L}_{MCS
+M}(A,N_f\to\infty) + \frac{m^2}2 f^{\mu}f_{\mu} - \frac{e^2}4 F_{\mu\nu}(f) g(\Box)
F^{\mu\nu}(f) \label{decoupled4} \ee

\no Once again the particle content of (\ref{decoupled3}) and (\ref{decoupled4}) must
be the same. Since the pure Chern-Simons term in (\ref{decoupled3}) contains only one
massless non-propagating mode we expect that the poles of ${\cal L}_{SD
+M}(f,N_f\to\infty)$ should reappear in (\ref{decoupled4}). We have already checked
that the spectrum of ${\cal L}_{SD +M}(f,N_f\to\infty)$ and ${\cal L}_{MCS
+M}(A,N_f\to\infty)$ are the same. However, the quadratic terms in the $f_{\mu}$ field
in (\ref{decoupled4}) contain apparently a new pole of the type $1/(a_0 - a_2 k^2)$
with $a_0,a_2$ given in (\ref{aaa}). This pole was not present in (\ref{decoupled3})
but it turns out that the propagator of the ${\cal L}_{MCS +M}(A,N_f\to\infty)$ model
has exactly a zero, see gauge invariant part of (\ref{pgmcs}), at the point $a_0 - a_2
k^2=0$ such that the poles of (\ref{decoupled3}) and (\ref{decoupled4}) still match as
expected. After the above discussions it becomes clear that if we replace the master
action given in (\ref{master1}) by the new one below we will have the same spectrum
for both dual theories:

\be {\cal L}_{Master}^{new}  \, = \, a_0 f^{\mu}f_{\mu} + a_1
\epsilon_{\alpha\beta\gamma}f^{\alpha}\partial^{\beta}f^{\gamma} + \frac{a_2}2
F_{\mu\nu}(f)F^{\mu\nu}(f) - a_1 \epsilon_{\alpha\beta\gamma}(A^{\alpha}-
f^{\alpha})\partial^{\beta}(A^{\gamma}-f^{\gamma}) \label{mastern} \ee

\no Performing a translation $A_{\mu} \to A_{\mu} + f_{\mu}$ we have a pure
Chern-Simons term plus the GSD model given in (\ref{gsd}) while a translation in the
$f_{\mu}$ field would lead to

\be {\cal L}_{Master}^{new}  \, \to \, {\cal L}_{NL-MCS} + f^{\mu}\left(a_0 g_{\mu\nu}
- a_2\Box \theta_{\mu\nu}\right)f^{\nu}\label{decoupled5}\ee \no where the new action
is given by:

\be {\cal L}_{NL-MCS}  \, = \, - a_1 \epsilon_
{\mu\nu\alpha}A^{\mu}\partial^{\nu}A^{\alpha} - F^{\mu\nu}(A)\frac{a_1^2}{2(a_0 - a_2
\Box)}F_{\mu\nu}(A) \label{nlgmcs} \ee

\no Notice that for $a_2=0$ we recover the duality between the SD and MCS models. The
corresponding propagator is the same as (\ref{pgmcs}). Thus, we do not need to check
that  the spectrum of ${\cal L}_{NL-HD-MCS}$ and ${\cal L}_{GSD}$ is the same, that
is, two physical massive particles without ghosts. Now let us check the dual map. The
new master action furnishes the equations of motion:

\beq d(A-f) \, &=& 0 \label{mneq1}\\
f_{\mu} \, &=& \, \tilde{f}_{\mu} \label{mneq2} \eeq

\no The duality transformation is defined as in section 2. The equation (\ref{mneq2})
is the same of the GSD model. From (\ref{mneq1}) and (\ref{mneq2}) we have $A_{\mu} +
\partial_{\mu}\phi \, = \, \tilde{A}_{\mu} $ which leads to $a_2^2 \left( \Box +
m_+^2\right)\left( \Box + m_-^2\right){\tilde A}_{\mu}  = 0$. Since the dual of a
total divergence vanishes we have $\tilde{A}_{\mu} = \tilde{\tilde{A}}_{\mu}$ which
guarantees the classical duality between ${\cal L}_{NL-MCS}$ and ${\cal L}_{GSD}$
using the map $f_{\mu} \leftrightarrow \tilde{A}_{\mu}$. On the other hand, from
(\ref{master1}) we have:

\beq \tilde{f}_{\mu} \ \, &=& \tilde{A}_{\mu} \label{meq1}\\
f_{\mu} \, &=& \, \tilde{f}_{\mu} \label{meq2} \eeq

\no They also lead to $\tilde{A}_{\mu} = \tilde{\tilde{A}}_{\mu}$. If $a_2=0 $,
 ${\cal L}_{Master}^{new}$ and ${\cal L}_{Master}$ are equivalent as (\ref{mneq1})
 and (\ref{meq1}). In the following we confirm that the dual equivalence persists at
 quantum level up to contact terms. If we
 define the generating function:

\be {\cal Z}(J) \, = \, \int {\cal D}A{\cal D}f \, e^{i\int d^3 x \left\lbrack {\cal
L}_{Master}^{New}(f,A) + \frac{\lambda(\partial_{\mu}A^{\mu})^2}{2} +
J^{\alpha}\tilde{A}_{\alpha}\right\rbrack} \label{zj1}\ee

\no After $A_{\mu} \to A_{\mu} + f_{\mu}$ and integrating over $A^{\mu}$ we obtain, up
to an overall constant,

\be {\cal Z}(J) \, = \, \int {\cal D}f \, e^{i\int d^3 x \left\lbrack {\cal
L}_{GSD}(f) + J^{\alpha}D_{\alpha\beta}J^{\beta} +
J^{\alpha}\tilde{f}_{\alpha}\right\rbrack} \label{zj2}\ee

\no with \be D_{\alpha\beta}=\frac{-1}{4 a_1 a_0^2}\left\lbrack[(a_2^2 \, \Box -
a_1^2)E_{\alpha\beta} - 2 a_1 a_2 \, \Box \, \theta_{\alpha\beta}
\right\rbrack\label{d}\ee

\no On the other side, if we integrate over $f_{\mu}$ in (\ref{zj1}) we have ${\cal
L}_{NL-HD-MCS} + \frac{\lambda(\partial_{\mu}A^{\mu})^2}{2} +
J^{\alpha}\tilde{A}_{\alpha}$. Therefore, deriving with respect to $J_{\alpha}$ we
deduce:

\be \left\langle \tilde{f}_{\alpha}(x) \tilde{f}_{\beta}(y) \right\rangle_{GSD} \, =
\, \left\langle \tilde{A}_{\alpha}(x) \tilde{A}_{\beta}(y) \right\rangle_{NL-MCS} \, -
\, D_{\alpha\beta} \delta^{(3)}(x-y) \label{cf1} \ee

\no Furthermore, if we define the generating function

\be {\cal Z}_{GSD}(J,j) \, = \, \int {\cal D}f \, e^{i\int d^3 x \left({\cal
L}_{GSD}(f)
 + J^{\alpha}\tilde{f}_{\alpha} + j^{\alpha}f_{\alpha} \right)} \label{zj3}\ee

\no and make $f_{\mu} \to f_{\mu} + J_{\mu}/2 a_0$ we have

\be {\cal Z}_{GSD}(J,j)  =  \int {\cal D}f \, \exp\left\lbrace i\int d^3 x
\left\lbrack {\cal L}_{GSD}(f)
 + (j^{\alpha} + J^{\alpha})f_{\alpha}
 - \frac{J_{\nu}j^{\nu}}{2a_0} - \frac{J_{\mu}H^{\mu\beta}J_{\beta} + J_{\nu}J^{\nu}}{4 a_0}
 \right\rbrack \right\rbrace \label{zj4}\ee

\no Where $H_{\alpha\beta}$ is the operator on the right handed side of (\ref{sdeq}),
i.e., $\tilde{f}_{\alpha} \equiv H_{\alpha\beta}f^{\beta}$. From (\ref{zj3}) and
(\ref{zj4}), see also \cite{Banerjee2,Bazeia2}, we have

\beq \left\langle f_{\alpha}(x) \tilde{f}_{\beta}(y) \right\rangle_{GSD} \, &=& \,
\left\langle f_{\alpha}(x) f_{\beta}(y) \right\rangle_{GSD} + \frac{g_{\alpha\beta}}{2
a_0} \delta^{(3)}(x-y) \label{cf2} \\
\left\langle \tilde{f}_{\alpha}(x) \tilde{f}_{\beta}(y) \right\rangle_{GSD} \, &=& \,
\left\langle f_{\alpha}(x) f_{\beta}(y) \right\rangle_{GSD} +
\frac{\left(g_{\alpha\beta}+ H_{\alpha\beta}\right) }{2 a_0} \delta^{(3)}(x-y)
\label{cf3} \eeq

\no  Since the GSD model is quadratic the relations (\ref{cf2}) and (\ref{cf3}) assure
that the classical self-duality $f_{\mu}=\tilde{f}_{\mu}$ holds also at quantum level
up to contact terms. Combining (\ref{cf1}) and (\ref{cf3}) we derive:

\be \left\langle f_{\alpha}(x) f_{\beta}(y) \right\rangle_{GSD} \, = \, \left\langle
\tilde{A}_{\alpha}(x) \tilde{A}_{\beta}(y) \right\rangle_{NL-MCS} \, - \,
\left(D_{\alpha\beta}- \frac{(g_{\alpha\beta}+ H_{\alpha\beta} }{2
a_0}\right)\delta^{(3)}(x-y) \label{cf4} \ee

\no We conclude that  ${\cal L}_{NL-MCS}(A)$ is dual to ${\cal L}_{GSD}(f)$ at
classical and quantum level (up to contact terms) with the map $f_{\mu}
\leftrightarrow \tilde{A}_{\mu}$. From this point of view ${\cal L}_{NL-MCS}(A)$ is on
the same footing of ${\cal L}_{HD-MCS}(A)$ but it has the advantage of having the same
spectrum of the GSD model and being ghost free.

We end up this section with three remarks. First, it is tempting to formally redefine
the gauge field $A_{\mu} \to \left(a_0 - a_2 \Box\right)^{1/2}A_{\mu}$ in order to
make ${\cal L}_{NL-MCS}(A)$ a local gauge theory. However, the higher derivative
theory thus obtained, although classically equivalent to the GSD model and simpler
than ${\cal L}_{HD-MCS}$, contains two poles $k^2=m_{\pm}^2$ and one of them is
necessarily a ghost now. This is not totally surprising, the situation is similar to
the massless Schwinger model in $1+1$ dimensions where the integration over the
fermionic massless fields gives rise to the effective (exact) gauge invariant action:

\be {\cal L} \, = \, - \frac 14 F_{\mu\nu}^2 - \frac {e^2}{4\pi} F^{\mu\nu}\frac
1{\Box}F_{\mu\nu} \label{l} \ee

\no The photon propagator from the above Lagrangian contains only one massive pole at
$k^2=e^2/\pi$ and no ghosts. However, if we try a naive redefinition $A_{\mu} \to
\sqrt{\pm \Box} A_{\mu}$ the action becomes local but the propagator will get a factor
$\pm\left\lbrack k^2(k^2-e^2/\pi)\right\rbrack^{-1}$. Whatever sign we choose we
always have a ghost field either at $k^2=0$ or $k^2=e^2/\pi$. In fact in this specific
case we can achieve a consistent local formulation by  introducing a scalar field
through $A_{\mu} = (\epsilon_{\mu\nu}\partial^{\nu}/\sqrt{\Box})\phi $. However, a
consistent formulation in terms of a vector gauge field can only be non-local to the
best we know.

The second remark concerns the uniqueness of the new master action. Clearly, the
important point in our proposal is that the gauge and the non-gauge fields are mixed
though a non-propagating term, in this case of $d=2+1$ we have used the Chern-Simons
term. One might try to generalize the new master action by using an arbitrary constant
in front of the mixed Chern-Simons term in (\ref{mastern}) instead of $a_1$. We have
done that and checked that all important conclusions about dual equivalence, spectrum
match, absence of ghosts, etc, hold for arbitrary values of this parameter but the
simplest dual gauge theory is obtained precisely for the theory suggested here.

At last, we remark that the duality between ${\cal L}_{NL-MCS}(A)$ and the GSD model
obtained by the new master action is also valid for the case where $a_0=a_0(\Box )\,
,\, a_2=a_2(\Box)$ are not constants. If the coefficient of the Chern-Simons mixing
term $a_1$ is kept constant we expect once again a spectrum equivalence otherwise the
match of the spectrum between the non-gauge and the gauge theory will no longer be
true in general.

\section{Conclusion}

Recently, the Noether gauge embedding and the master action procedures were applied in
different theories and different dimensions. In particular, very recently, one has
been trying to generalize them to nonabelian theories \cite{nonabelian} and models
defined on the non-commutative plane \cite{NC}. This embedding procedure usually
reproduces the result of a master action which is very convenient at quantum level.
Here we have shown how we can modify this master action to avoid extra poles in the
propagator and assure that both dual theories have the same spectrum. Though, we have
used a generalized self-dual model as an example, it is clear that similar ideas can
be useful whenever we make use of interpolating master actions. The key point is to
choose a non-propagating term to mix the gauge and non-gauge fields. To lend support
to our suggestion we have compared the spectrum of the effective action generated by
the minimal coupling of fermions (parity invariant) to the self-dual field and the
effective action generated by its dual gauge theory, both in the limit $N_f \to
\infty$ where the quadratic effective actions become exact. As a by product we have
shown that those theories contain two massive poles in their spectrum and are free of
ghosts and tachyons\footnote{Clearly, a more general matter coupling, including even a
non-minimal coupling with scalar fields also deserves to be examined but that is
beyond the aims of this letter.}. The price we have paid for a ghost free dual gauge
theory was locality. However, it is well know that it is not easy to have massive
vector fields and still keep local gauge invariance, sometimes we have to break
locality as in the massless Schwinger model. Since the GSD model contains two massive
vector modes, one might speculate that the usual topological mechanism \cite{DJT}  to
generate mass for a gauge field in $d=2+1$, with the help of higher derivatives, could
not cope with two massive poles without generating unwanted extra poles. More work is
necessary to clarify this point.

\section{Acknowledgements}

This work was partially supported by \textbf{CNPq}, a Brazilian research agency. We
thank Alvaro de Souza Dutra for discussions and bringing reference \cite{DN} to our
attention. A discussion with Marcelo Hott is also acknowledged.

\end{document}